\documentclass[12pt]{article}
\usepackage[T1]{fontenc}
\usepackage{graphicx,authblk,hyperref,url,parskip,array}
\pdfoutput=1

\author[1]{Alexey Shipunov}

\affil[1]{Minot State University, Minot ND, 58707}

\title{
Ripeline and Rmanual speed up\\
biological research and reporting
}

\date{}

\newenvironment{Hang}{\leftskip2em\parindent-2em\parskip1ex\mbox{}}{}

\begin{document}

\maketitle

\begin{abstract}

The emergence of R, a freely available data analysis environment, brought to the researcher in any science field a set of well-concerted instruments of immense power and low cost. In botany and zoology, these instruments could be used, for example, to speed up work in two distant but related fields: analysis of DNA markers and preparation of natural history manuals. Both of these tasks require a significant amount of monotonous work, which could be automated with software.

I developed ``Ripeline and ``Rmanual,'' two highly customizable R-based applications, designed with a goal of simplicity, reproducibility, and effectiveness. Ripeline is a pipeline that allows for a continuously updated analysis of multiple DNA markers. Rmanual is a ``living book'' which allows the creation and continuous update of manuals and checklists.

Comparing with more traditional ways of DNA marker analysis and manual preparation, Ripeline and Rmanual allow for a significant reduction of time, which is usually spent doing repetitive tasks. They also provide tools which can be used in a broad spectrum of further applications.

\end{abstract}


\section{Introduction}

DNA extraction, PCR amplification, and sequencing are now widely available for researchers, and many computer software exists to generate phylogenetic trees from sequences. However, the amount of work required is often overwhelming. Even more complicated is the situation when the project requires an update, and all steps have to be repeated. This is where software might be able to help. It must link sequence generation and phylogeny estimation in the form of a custom pipeline that works with multiple fragment types. Also, it should be flexible (new analyses are easy to add), contemporary, updatable (allows to rebuild the trees every time databases change), and use both local and globally available data. It is also hard to find an existing portable software which allows for proper sequence concatenation and gap coding. Ideally, this pipeline should also allow to check, for example, which species within a group are not yet successfully sequenced, or whether species names are congruent between taxonomic and sequence parts of data.

Checklists and natural history manuals are cornerstones of biodiversity research. However, much of biodiversity-related literature has spent years in preparation. When taxonomic databases contain numerous name entries, descriptions, habitat notes, and image links, the task to produce a checklist, illustrated checklist, or manual is also overwhelmingly complex, especially in the case when the product needs an update. The preparation process includes a large amount of technical work, for example, adding images and descriptions to species. I believe that a semi-automatic approach that combines processing of relational text tables and typographic production will allow for a significant increase in speed, and therefore will improve the production process of manuals, checklists, pamphlets, and other similar literature.

Applications presented below help to improve both sequence analysis and manual preparation. While they are devoted to different tasks, they have much in common. Both are (1) based on ``flat'' text tables, (2) use R as a backbone for databasing and organizing the input and output from the external software, and (3) designed with the idea of continuous updates based on positive feedback. The essential part of both applications are R scripts. Started as a statistical environment (R Core Team, 2019), R is now rapidly gaining popularity as the general data processing language for science, especially in biology. This popularity is mostly due to extensive documentation, free programming style, many high-level routines, advanced data manipulation and plotting abilities. R is easy to embed, integrate, and call (both from it and out of it). R scripts are one of the simplest ways to deploy and reproduce the data processing workflow. As Ripeline and Rmanual are based on reproducibility, portability, and integration, their R backbone is indispensable.

\section{Methods and Results}

\subsection{Ripeline (R-based sequence analysis pipeline)}

\begin{figure}[ht]\centering
\includegraphics[width=.65\textwidth]{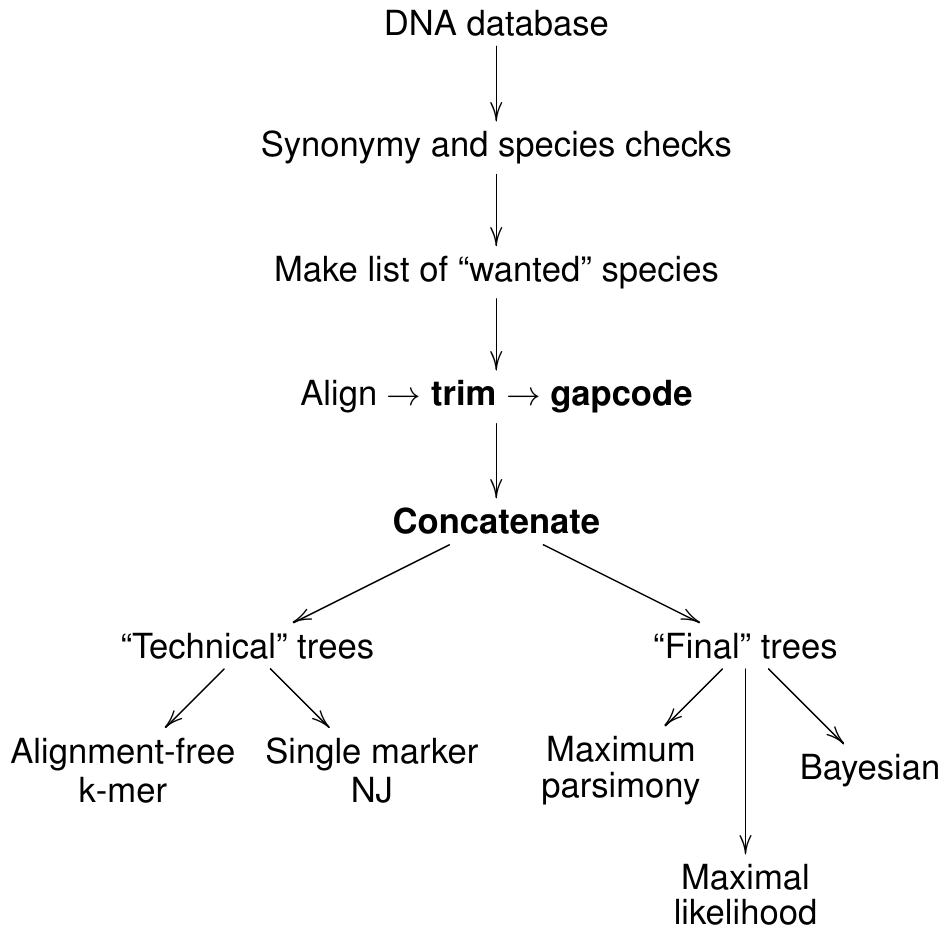}
\caption{
Ripeline: the main pipeline.
}\label{1}
\end{figure}

Ripeline is an R-based pipeline application that allows for the wide variety of sequence analyses (Fig.~\ref{1}). In the core, it is a set of R and UNIX shell script templates (i.e., simple text files), which are extensively commented and easy to adapt for the needs of a particular project. I made the Ripeline when I realized that I must run four independent phylogenetic projects, each with plenty of marker data, and with only the restricted help from undergraduate students. It saved a significant amount of my time, which otherwise would have been spent on repetitive tasks. Now, thanks to Ripeline, the first results of these projects are already published (Shipunov et al., 2019).

Ripeline includes both R-based and standalone methods of analysis. While many of these are available separately, the actual workflows are diverse, not standardized, and therefore not fully reproducible. Ripeline allows for the fully standardized, reproducible sequence analyses. Installation of Ripeline is also completely portable, as it is equal to downloading the single directory.

Each Ripeline script (Table 1) is numbered (like in the UNIX ``init'' startup system) and works independently (so if it fails for any reason, others will continue to work). It also outputs a full report of what was done, thanks to the ``Rresults'' utility from ``shipunov'' package (Shipunov, 2019).

Ripeline starts by checking the local DNA database (Table~\ref{4}, Fig~\ref{1}).

\begin{table}[ht]\centering
\begin{tabular}{|>{\tt}l|>{\raggedright\arraybackslash}p{6.5cm}|}\hline
03\_checks.r &
checks input tables \\\hline
04\_duplicated\_ids\_check.r &
checks duplicated IDs \\\hline
13\_make\_wanted.r &
outputs names of non-sequenced species \\\hline
20\_make\_sets.r &
combines DNA data into ``sets'', one for each marker \\\hline
30\_align.r &
run the external multiple alignment tool \\\hline
31\_trim.r &
trims flanks of alignments \\\hline
32\_gapcode.r &
gap coding \\\hline
40\_concatenate\_and\_stat.r &
concatenates sequences and outputs statistics \\\hline
51\_make\_r\_raw\_kmer\_trees.r &
builds ``raw'' (based on all data) k-mer trees \\\hline
52\_make\_r\_semistrict\_kmer\_tree.r &
builds k-mer trees on semistrict super-matrix \\\hline
53\_make\_r\_nj\_single\_marker\_trees.r &
builds NJ (neigbor joining) single marker trees \\\hline
61\_make\_r\_mp\_semistrict\_tree.r &
builds MP (maximum parsimony) trees \\\hline
71\_make\_r\_ml\_modeltest.r &
tests maximal likelihood models \\\hline
72\_make\_r\_ml\_trees.r &
estimates ML (maximal likelihood) trees using internal R functions \\\hline
73\_make\_raxml\_trees.r &
estimates ML trees with RAxML \\\hline
81\_make\_mrbayes\_semistrict\_tree.r &
estimates Bayesian trees with MrBayes \\\hline
\end{tabular}
\caption{
Ripeline: basic R scripts
}\label{4}
\end{table}

This database is the tab-delimited file with the following fields:

\begin{description}

\item[\texttt{SOURCE}] (where from). By default, Ripeline trusts locally obtained sequences more.

\item[\texttt{SEQUENCE.ID}] Typically, something like ``K-216'' where ``K'' is a shortcut for the project.

\item[\texttt{FRAGMENT}] Fragment name, like ``abcd'' or ``efgh''.

\item[\texttt{SPECIES.NEW}] Species name.

\item[\texttt{COMMENT}]

\item[\texttt{SELECT}] (0 or 1). Which sequences to use.

\item[\texttt{SEQUENCE}] The actual DNA string; allowed letters are ``ATGCN''.

\end{description}

DNA-related data combined into ``sets'' (one per marker), aligned, trimmed, gap coded, and then concatenated in super-matrices. Two types of concatenation are employed:

\begin{description}

\item[Strict concatenation] Based on sequence IDs of locally obtained sequences.

\item[Semi-strict concatenation] The next step. It uses ``strict'' dataset and adds sequences of any origin in order to fill all gaps.

\end{description}

Ripeline uses some external tools like MUSCLE (Edgar, 2004), RAxML (Stamatakis, 2014), and MrBayes (Ronquist \& Huelsenbeck, 2003), but can also work without them. Some proficiency in R is required to adopt these scripts for the needs of a particular project. However, nothing beyond small modifications (e.g., changing the size of the output image file) is necessary. Only rudimentary shell scripting knowledge is required, on the level of commenting or uncommenting particular lines. All in all, I believe that Ripeline is suitable for users without excessive programming skills.

As all its components are cross-platform, Ripeline works on all major platforms. Ripeline is fast. Template scripts are built around an artificial example which involves two DNA markers from 12 species, multiple alignment, trimming flanks, gap coding, concatenation, creation of ``technical'' k-mer and NJ trees, a maximum parsimony example, maximum likelihood (ML) hypotheses testing, two examples of ML, and Bayesian tree estimation. With parameters set to minimum (like 100 bootstrap replicates), the whole pipeline takes about 2 minutes on an average laptop.

Ripeline is the new software, but there is already a series of use cases. Results of the first project are recently published (Shipunov et al., 2019), and two other projects are accessible in the author’s open repository (\url{http://ashipunov.info/shipunov/open/}).

\subsection{Rmanual (R-based natural history manual)}

Software that assists in producing the natural history manual should be able to extract data from databases, use images, and output typographically formatted text. To my knowledge, R (R Core Team, 2019) and \TeX\ (Knuth, 1984) are the best candidates for making this kind of software.

Rmanual requires working \TeX\ and R installations, plus one additional R package. All of these are available on all major platforms. Rmanual includes only one R script and also one shell script whose job is to use the former, and then run the \TeX\ engine. R code is short, fully commented, and is easy to modify. There are only a few \TeX\ definitions, which are easy to understand and modify.

\begin{figure}[ht]\centering
\includegraphics[width=.98\textwidth]{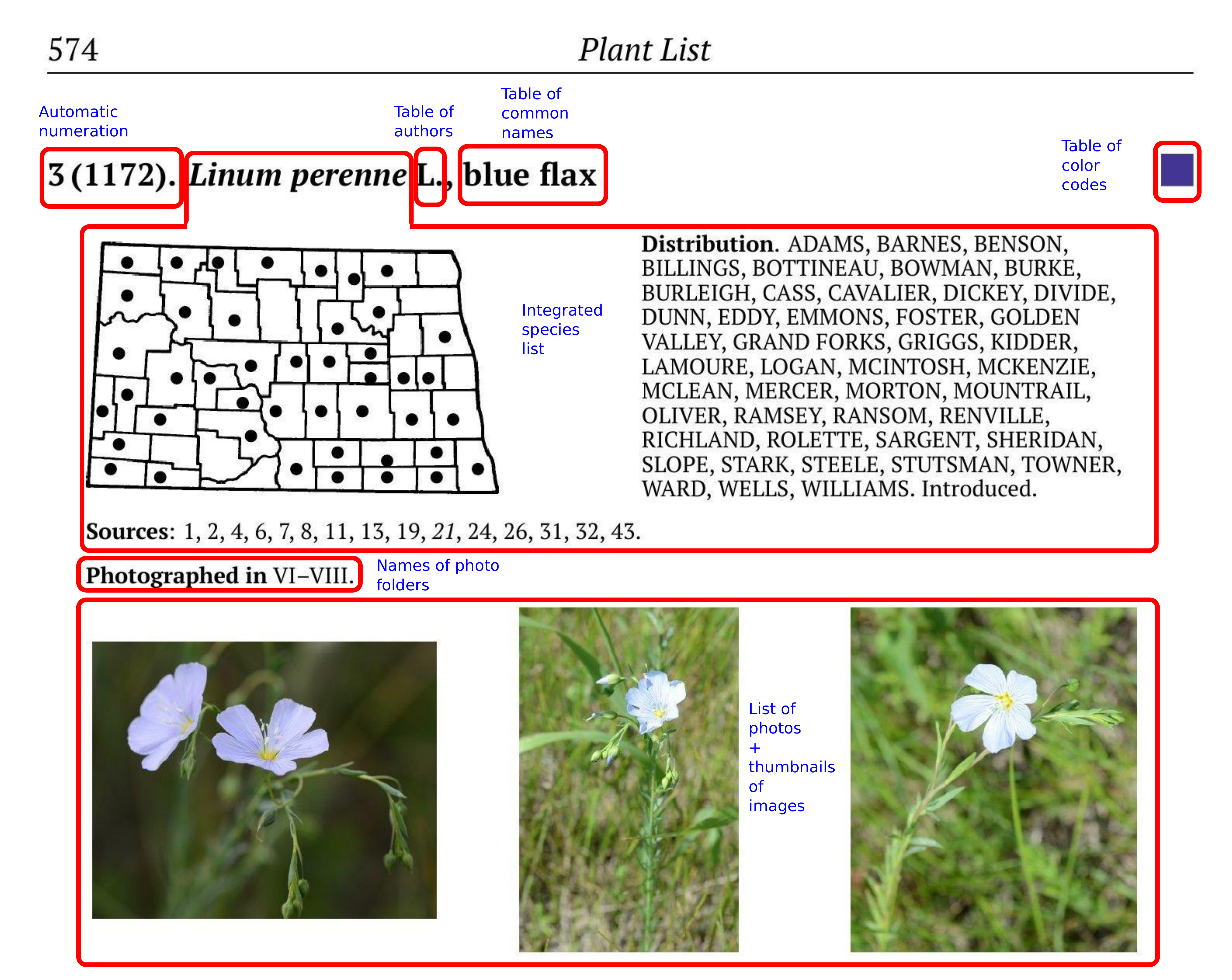}
\caption{
Rmanual: the structure of the output manual page.
}\label{2}
\end{figure}

Before starting the Ripeline, I recommend utilities like TNRS (Boyle et al., 2013) for the data cleaning and normalization. The basic idea of the Rmanual is that after R outputs a text table (made from inter-combined input tables), \TeX\ uses it as a ``body'' (main, taxonomic part) of the book. In the end, each page inside the PDF book consists of multiple table pieces (Fig.~\ref{2}).

One of these pieces might look like:

\begin{verbatim}
\FF{Subgenus \KK{Minores}} 
 \SP{\KK{Kubrickus heus}}
 \DD{Photosynthetic, non-carnivorous, motile}
 \II{\I{kubrick_h.pdf}}
\end{verbatim}

In the piece above, each field of the output table is formatted as an argument of some \TeX\ command: \verb|\FF{}| for higher categories, \verb|\SP{}| for species names, \verb|\DD{}| for species descriptions, and \verb|\II{}| for figures.

\begin{figure}[ht]\centering
\includegraphics[width=.9\textwidth]{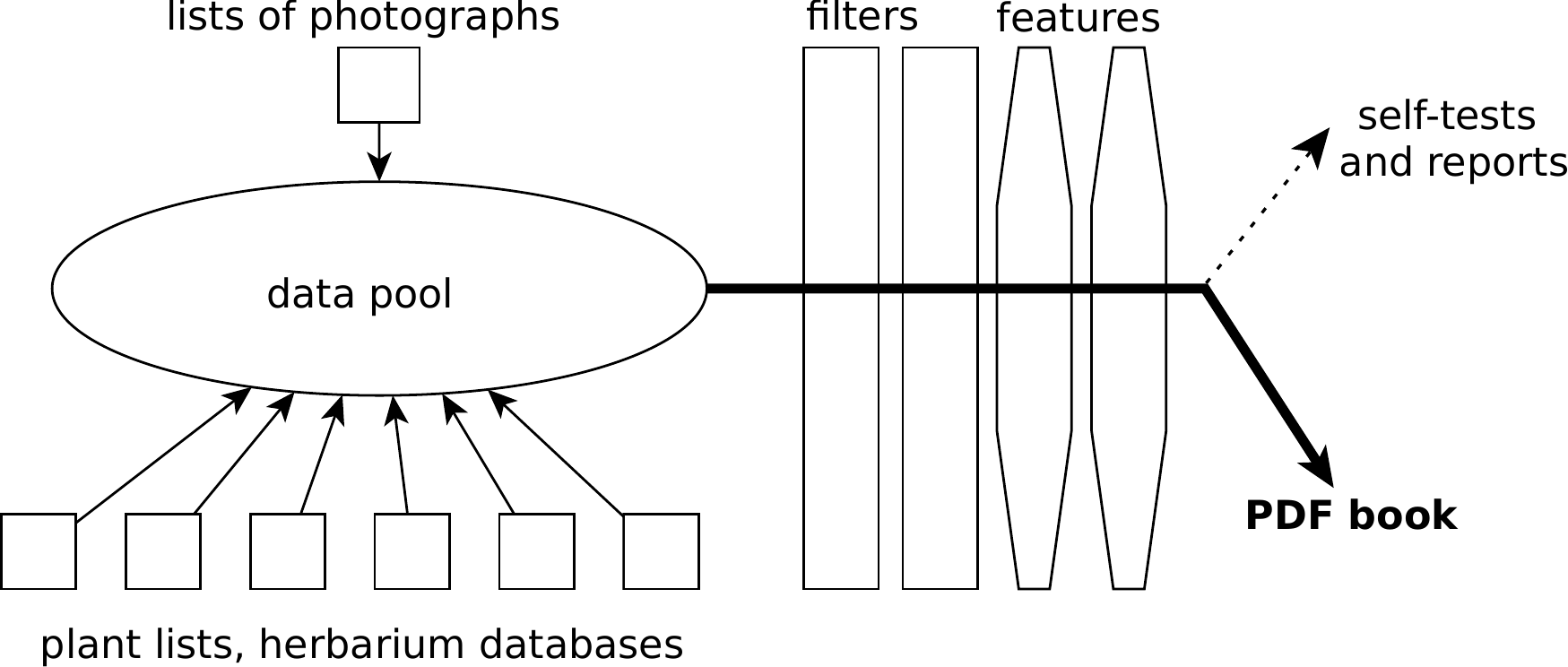}
\caption{
Rmanual: the main pipeline.
}\label{3}
\end{figure}

Inter-combining of input tables uses the initial species list, ``filters'' and ``features''. Filters (like a table connecting synonyms and accepted names) convert data, whereas features (like a table connecting genera and families) add new information (Fig.~\ref{3}).

The header and footer of the book (the title page, first and last pages) should be prepared manually using the supplied template. Apart from a PDF book, Rmanual also outputs diagnostic data (e.g., which species do not correspond with images).

The main strength of Rmanual is, therefore, the production of semi-automatic, typographically ready output from the set of ``flat'' text tables and images. Besides, Rmanual allows for the constant update of the output and therefore produces books that are not only semi-automatic but also ``living.''

Rmanual, similarly to Ripeline, requires some knowledge in R and \TeX\ but does not require extensive skills. If the researcher can modify R and \TeX\ files (note again that they are simple text files), it should be enough to run Rmanual with their data and make simple PDF books. I tried the Rmanual approach several times (Shipunov, 2013; Shipunov et al., 2015; Shipunov, 2018), and each time the process of adaptation did not take more than a few hours.

Rmanual is very fast. When preparing this paper, I made a new illustrated checklist, which included 122 plant species and their images within just two hours. The actual book production (creation of a typographically ready PDF file) takes only a few seconds on an average laptop. Even if an inexperienced person will be ten times slower, 20 hours for a small book is a good result. Therefore, I believe that when all data is ready, with a help from Rmanual or with a Rmanual-like approach, the production of typographic results might span only a few days.

The illustrated checklist of North Dakota plants (Butler, 2013; Shipunov, 2018) is one of the most advanced use cases of Rmanual. Two other working examples of Rmanual, which use real data, are presented on Github: \url{https://github.com/ashipunov}.

\section{Conclusions}

The biggest drawback of Ripeline and Rmanual is that the user must know some basics of R and (in case of Rmanual) \TeX. However, both of these systems are free, and more importantly, extensively documented and community-supported. Free manuals can be accessed both on CTAN (e.g., \url{https://ctan.org/pkg/lshort-english} and \url{https://ctan.org/pkg/beginlatex}) and CRAN (\url{https://cran.r-project.org/other-docs.html}.)

There are only a few tools comparable with Ripeline. PhyScraper (McTavish, 2019), for example, is the Python-based pipeline intended to do a similar job. However, it is still in development (some parts are therefore not available), and also uses a virtual environment which brings extra abstraction layers. Phylogeny.fr (Dereeper et al., 2008) implements a similar idea but is Web-oriented (i.e., depends on network services which can slow down the whole process) and restricts the amount of allowable data. Other similar tools like phyloGenerator (Pearse \& Purvis, 2013) or ReproPhylo (Szitenberg et al., 2015) are intended mostly for publicly available (GenBank) sequences.

The R package monographaR (Reginato, 2016) works on principles similar to Rmanual, but its goal is taxonomic monographs and not checklists or manuals (it is, therefore, more complicated than Rmanual), it also does not involve \TeX\ which consequently lowers the typographic quality of the result.

Ripeline, Rmanual, and two additional example manuals are freely available on my GitHub: \url{https://github.com/ashipunov}. Ripeline and Rmanual are also available as appendices to my ``Visual Statistics. Use R!'' book: \url{http://ashipunov.info/shipunov/software/r/r-en.htm}.

\section{Acknowledgements}

The author is grateful to the Vitalij Arnold, Sergej Sukhov, Vera Shipunova, and authors of R and \TeX\ software.

\section{Literature cited}

\begin{Hang}

Boyle, B., N. Hopkins, Z. Lu, J. A. R. Garay, D. Mozzherin, T. Rees, N. Matasci, et al. 2013. The taxonomic name resolution service: an online tool for automated standardization of plant names. BMC Bioinformatics. 14: 16.

Butler, J.L., 2013. Flora of North Dakota: Checklist by Alexey Shipunov. Native Plants Journal, 14(3), pp.267-268.

Dereeper, A., V. Guignon, G. Blanc, S. Audic, S. Buffet, F. Chevenet, J. F. Dufayard, et al. 2008. Phylogeny.fr: robust phylogenetic analysis for the non-specialist. Nucleic Acids Research. 36: 465–469.

Edgar, R.C. 2004. MUSCLE: multiple sequence alignment with high accuracy and high throughput. Nucleic Acids Research 32: 1792–1797.

Huelsenbeck, J.P., \& Ronquist, R. 2001. MrBayes: Bayesian inference of phylogenetic trees. Bioinformatics 17: 754–755.

Knuth, D.E. 1984. The \TeX book. Reading: Addison-Wesley.

McTavish, E.J. physcraper. URL:\\ \url{https://github.com/McTavishLab/physcraper} [last accessed 28 Nov. 2019]

Pearse, W.D. and Purvis, A. 2013. phyloGenerator: an automated phylogeny generation tool for ecologists. Methods in Ecology and Evolution. 4: 692–698.

R Core Team. 2019. R: A language and environment for statistical computing. R Foundation for Statistical Computing, Vienna, Austria. URL: \url{http://www.R-project.org/} [last accessed 19 Nov. 2019]

Reginato, M. 2016. monographaR: an R package to facilitate the production of plant taxonomic monographs. Brittonia. 68: 212–216.

Shipunov, A. 2013. Atlas of plants of Tver region. 2013. URL: \url{http://ashipunov.info/shipunov/moldino/nauka/atlas.pdf} [last accessed 19 Nov. 2019]

Shipunov, A. 2018. Flora of North Dakota: Illustrated checklist. Version 2. Ed.: Kartesz, J., and Nishino, M. ISBN 978--1-4675--6379--6. URL: \url{http://ashipunov.info/shipunov/fnddb/index.htm} [last accessed 19 Nov. 2019]

Shipunov, A. 2019. shipunov: Miscellaneous Functions from Alexey Shipunov. R package version 1.2. URL: \url{https://CRAN.R-project.org/package=shipunov} [last accessed 19 Nov. 2019]

Shipunov, A., Gladkova S., Timoshina P., Lee H. J., Choi J., Despiegelaere S., and Connolly B. 2019. Mysterious  chokeberries: new data on the diversity and phylogeny of Aronia Medik. (Rosaceae). European Journal of Taxonomy 570: 1–14.

Shipunov, A., L. Abramova, J. Beaudoin, J. H. Choi, D. Fry, R. Perry, V. Shipunova, J. Schumaier, and J. Theodore. 2015. Flora of North Dakota project. Phytoneuron 16: 1–10.

Stamatakis, A. 2014. RAxML version 8: a tool for phylogenetic analysis and post-analysis of large phylogenies. Bioinformatics. 30:1312–1313.

Szitenberg, A., M. John, M. L. Blaxter, and D. H. Lunt. 2015. ReproPhylo: an environment for reproducible phylogenomics. PLoS Computational Biology. 11:1004447.

\end{Hang}

\end{document}